\documentclass{ws-procs9x6}
\usepackage{rsfs}

\begin{document}

\title{A GAUGE-INVARIANT MECHANISM \\
FOR QUARK CONFINEMENT\\
AND A NEW APPROACH TO THE MASS GAP PROBLEM}

\author{KEI-ICHI KONDO}

\address{Department of Physics, Faculty of Science, 
Chiba University, \\
Chiba 263-8522, Japan\\
E-mail: kondok@faculty.chiba-u.jp
}

\begin{abstract}
We give a gauge-invariant description of the dual superconductivity for deriving quark confinement and mass gap in Yang-Mills theory.  
\end{abstract}


\bodymatter

\section{Introduction}

The fundamental degrees of freedom of QCD, i.e., {quarks} and {gluons}, have never been observed in experiments. 
Only  the color singlet combinations, {hadrons (mesons and baryons)} and {glueballs}, are expected to be observed.
We wish to answer the question why color confinement occurs. 
In particular, we wish to clarify what the mechanism for quark confinement is.

Dual superconductivity picture 
proposed by Nambu, 't Hooft and Mandelstam\cite{dualsuper} in 1970s is based on the electric--magnetic duality of the ordinary superconductivity.  
In order to apply this idea to describe the dual superconductivity in Yang-Mills theory, however, we must answer the questions:
\begin{enumerate}
\item 
How to extract the {``Abelian'' part} responsible for quark confinement  from the non-Abelian gauge theory in the {gauge-invariant way}  without losing characteristic features of non-Abelian gauge theory, e.g., asymptotic freedom.\cite{KondoI}

\item  
How to define the {magnetic monopole} to be condensed in  Yang-Mills theory in the {gauge-invariant way} even in absence of any  fundamental scalar field, in sharp contrast to the Georgi-Glashow model.

\end{enumerate}
In this direction, a crucial idea called the Abelian projection was proposed by 't Hooft\cite{tHooft81}.
Recall that the Wilson criterion of quark confinement is the area decay law of the Wilson loop average:
\begin{align}
\Big\langle 
{\rm tr} \left[ \mathscr{P} \exp \left\{ ig \oint_{C} dx^\mu \mathscr{A}_\mu(x) \right\} \right]
\Big\rangle_{YM} \sim e^{-\sigma_{NA} |S|} .
\end{align}
Recent investigations have shown that quark confinement based on the dual superconductor picture is realized in   
the maximal Abelian gauge (MAG)\cite{KLSW87,KondoII}: 
the continuum form of MAG for SU(2) is (a  background gauge)  given by
\begin{equation}
  [\partial_\mu \delta^{ab} - g \epsilon^{ab3}A_\mu^3(x)] A_\mu^b(x) = 0 \ (a,b=1,2) ,
\end{equation}
for the Cartan decomposition:
$
\mathscr{A}_\mu=A_\mu^a \frac{\sigma^a}{2} + A_\mu^3 \frac{\sigma^3}{2} \ (a=1,2).
$
Numerical simulations on a lattice in  MAG yield surprisingly the area decay law for the Abelian(-projected) Wilson loop in Yang-Mills theory:
\begin{align}
  \Big\langle \exp i g \oint_{C} dx^\mu A_\mu^3(x) \Big\rangle_{YM}^{MAG} \sim e^{-\sigma_{Abel} |S|} ~~!
\end{align}
Remarkable results are the saturation of the string tension\cite{EI82}:
{Abelian dominance}\cite{SY90} $\Leftrightarrow$ $\sigma_{NA}  \sim \sigma_{Abel}$
and the {Monopole dominance}\cite{SNW94} $\Leftrightarrow$ $\sigma_{Abel}  \sim \sigma_{monopole}$ 
for the decomposition:
$
  A_\mu^3 = \text{Photon~part}+\text{Monopole part} .
$


However, we have still problems:

\noindent
$\bullet$ The Abelian projection and MAG break  SU(2) color symmetry  explicitly. 

\noindent
$\bullet$ Abelian dominance has never been observed in  gauge fixings other than MAG. 

Then one raises the question: the dual superconductivity might not be a gauge-invariant concept?
We wish to discuss how to cure these shortcomings.
In this talk, we argue that  
\begin{enumerate}
\item 
{[gauge-invariant ``Abelian''  projection]}
The {``Abelian'' part} $\mathbb{V}_\mu$ responsible for quark confinement can be extracted from the non-Abelian gauge theory
by using a nonlinear change of variables in the {gauge-invariant way}  without breaking color symmetry. 

\item  
{[infrared ``Abelian''  dominance]} 
The remaining part $\mathbb{X}_\mu$ acquires the mass to decouple in the low-energy region, leading to infrared ``Abelian'' $\mathbb{V}_\mu$ dominance. 
The dynamical mass originates from  the existence of gauge-invariant dimension-2 vacuum condensate $\left< \mathbb{X}_\mu^2 \right>\ne 0$. 
\end{enumerate}

\section{Non-Abelian Stokes theorem for the Wilson loop operator}

The Wilson loop operator for non-Abelian gauge field 
\begin{align}
  W_C[\mathscr{A}] 
  := {\rm tr} \left[ \mathscr{P} \exp \left\{ ig \oint_{C} dx^\mu \mathscr{A}_\mu(x) \right\} \right]/{\rm tr}({\bf 1})   ,
\end{align}
is rewritten into an Abelian form called the Diakonov-Petrov (DP) version of the non-Abelian Stokes theorem which was for the first time derived in ref.\cite{DP89} for SU(2).  This was rederived in refs.\cite{KondoIV,HU99} to be extented to SU(N) in refs.\cite{KT99,HU99}.  
Once a unit vector field $\bm{n}(x)$ is introduced, the SU(2) version is 
\begin{align}
  W_C[\mathscr{A}]  =& \int d\mu_S(\bm{n}) \exp \left\{ i g{J \over 2} \int_{S: \partial S=C} dS^{\mu\nu} G_{\mu\nu} \right\} , 
  \nonumber\\
  G_{\mu\nu}(x) =& \partial_\mu [\bm{n}(x) \cdot \mathscr{A}_\nu(x)] -  \partial_\nu [\bm{n}(x) \cdot \mathscr{A}_\mu(x)]   
  - g^{-1} \bm{n}(x) \cdot [\partial_\mu \bm{n}(x) \times \partial_\nu \bm{n}(x) ] ,
  \nonumber\\
     n^A(x) \sigma^A =& U^\dagger(x) \sigma^3 U(x) ,
     \quad \mathscr{A}_\mu(x)=\mathscr{A}_\mu^A(x) \sigma^A/2 ,
\end{align}
and $d\mu_S(\bm{n})$ is the product of the invariant measure on SU(2)/U(1) over $S$: 
\begin{align}
  d\mu_S(\bm{n})  :=\prod_{x \in S}d\mu(\bm{n}(x)) ,
  \quad
  d\mu(\bm{n}(x)) = {2J+1 \over 4\pi} \delta(\bm{n}(x) \cdot \bm{n}(x)-1) d^3 \bm{n}(x) .  
\end{align}
The ``Abelian'' field strength  $G_{\mu\nu}$ is SU(2) gauge invariant, since it is cast into the manifestly SU(2) invariant form:
\begin{align}
G_{\mu\nu}(x) =&  \bm{n}(x) \cdot \mathscr{F}_{\mu\nu}(x)  
  - g^{-1} \bm{n}(x) \cdot (D_\mu \bm{n}(x) \times D_\nu \bm{n}(x))
\nonumber\\
=& 2 {\rm tr} \left\{  \bm{n}(x) \mathscr{F}_{\mu\nu}(x)  
  + ig^{-1} \bm{n}(x)  [D_\mu \bm{n}(x), D_\nu \bm{n}(x) ] 
 \right\}  ,
\end{align}
where the gauge transformation is given by
$
     \bm{n}(x) \rightarrow U^\dagger(x) \bm{n}(x) U(x)   ,
$
$D_\mu \bm{n}(x) \rightarrow U^\dagger(x) D_\mu \bm{n}(x) U(x)   ,
$
$\mathscr{F}_{\mu\nu}(x) \rightarrow U^\dagger(x) \mathscr{F}_{\mu\nu}(x) U(x)   .
$
Note that 
$G_{\mu\nu}$ has the same form as the 't Hooft--Polyakov tensor under the identification
$
  n^A(x) \leftrightarrow \hat{\phi}^A(x):=\phi^A(x)/|\phi(x)| .
$
Therefore, the magnetic current $k_\mu$ obeying the 
topological conservation law $\partial^\mu k_\mu=0$ is defined by
\begin{align}
  k^\mu(x) :=& \partial_\nu {}^*G^{\mu\nu}(x) 
=   (1/2) \epsilon^{\mu\nu\rho\sigma}\partial_{\nu}
              G_{\rho\sigma}(x) .
\end{align}

\section{Reformulation of Yang-Mills theory based on the non-linear change of variables}

Using the Diakonov-Petrov NAST, 
we arrive at the expression for the Wilson loop average
$W(C):=\left< W_C[\mathscr{A}] \right>_{YM}$:
\begin{align}
W(C)  
=& \tilde{Z}_{\rm YM}^{-1} \int \mathcal{D}\mu[\bm{n}] 
 \int \mathcal{D}\mathscr{A}_\mu   e^{ iS_{\rm YM} } \tilde{W}_{\mathscr{A}}(C) 
= \frac{\int \mathcal{D}\mu[\bm{n}] \int \mathcal{D}\mathscr{A}_\mu  e^{ iS_{\rm YM} } \tilde{W}_{\mathscr{A}}(C) }{\int \mathcal{D}\mu[\bm{n}] \int \mathcal{D}\mathscr{A}_\mu   e^{ iS_{\rm YM} }},
\label{W1}
\end{align} 
where we have defined the {\it reduced} Wilson loop operator 
\begin{align}
 \tilde{W}_{\mathscr{A}}(C) 
 =  \exp \left\{ i g \frac{J}{2}  \int_{S:\partial S=C} d^2S^{\mu\nu} G_{\mu\nu} \right\}  ,
 \label{reducedW}
\end{align} 
and  the new partition function  
$
 \tilde{Z}_{{\rm YM}}  =\int \mathcal{D}\mu[\bm{n}] 
\int \mathcal{D}\mathscr{A}_\mu \exp (iS_{{\rm YM}}[\mathscr{A}]) . 
$
\\
Here we have inserted the unity into the functional integration, 
$
1=\int \mathcal{D}\mu[\bm{n}]
\equiv \int \mathcal{D}\bm{n}  \delta(\bm n\cdot\bm n-1)
:= \prod_{x \in \mathbb{R}^D} \int [d\bm{n}(x)]  \delta(\bm n(x) \cdot \bm n(x)-1) .
$

Suppose a unit vector field $\bm{n}(x)$ is given as a functional of $\mathscr{A}_\mu(x)$
\begin{align}
\bm{n}(x) = \bm{n}_{\mathscr{A}}(x) .
\end{align}
Then the ``decomposition'' (once known as    Cho-Faddeev-Niemi (CFN) decomposition.\cite{DG79,Cho80,FN98,Shabanov99})
\begin{align}
 \mathscr{A}_\mu(x) 
:=& \overbrace{\underbrace{c_\mu(x) \bm{n}(x)}_{\mathbb{C}_\mu:\text{restricted~potential}}
 +    \underbrace{ g^{-1}  \partial_\mu \bm{n}(x)  \times \bm{n}(x) }_{\mathbb{B}_\mu:\text{magnetic~potential}}}^{\mathbb{V}_\mu}  
 + \underbrace{ \mathbb{X}_\mu(x)}_{\text{covariant~potential}} ,
\end{align}
with
\begin{align}
c_\mu(x)
  ={\bm n}(x) \cdot \mathscr{A}_\mu(x), 
\quad 
\mathbb X_\mu(x)
  =g^{-1}{\bm n}(x)\times D_\mu[\mathscr{A}]{\bm n}(x) ,
\end{align}
is regarded as  {a non-linear change of variables} (NLCV): 
\begin{equation} 
(\mathscr{A}_\mu^A(x) \rightarrow) \quad 
{\bm n}^A(x), \mathscr{A}_\mu^A(x) \rightarrow {\bm n}^A(x), c_\mu(x), \mathbb{X}_\mu^A(x) .
\end{equation}

A remarkable property of NLCV is that the curvature tensor $\mathscr{F}_{\mu\nu}[\mathbb{V}]$ obtained from the connection $\mathbb{V}_\mu$
is parallel to $\bm{n}$ and its magnitude $G_{\mu\nu}$   coincides exactly with  $G_{\mu\nu}$ appearing in NAST of the Wilson loop operator: 
\begin{align}
 \mathscr{F}_{\mu\nu}[\mathbb{V}](x) 
:=& \partial_\mu \mathbb{V}_\nu(x)  - \partial_\nu \mathbb{V}_\mu(x)   + g \mathbb{V}_\mu(x)  \times \mathbb{V}_\nu(x) 
=  \bm{n}(x) G_{\mu\nu}(x) ,
\nonumber\\ 
\rightarrow G_{\mu\nu}(x) =& \bm{n}(x) \cdot \mathscr{F}_{\mu\nu}[\mathbb{V}](x)  
=  \partial_\mu [\bm{n}(x) \cdot \mathscr{A}_\nu(x)] -  \partial_\nu [\bm{n}(x) \cdot \mathscr{A}_\mu(x)]   
 \nonumber\\ 
& \quad\quad\quad\quad\quad\quad\quad\quad\quad - g^{-1} \bm{n}(x) \cdot [\partial_\mu \bm{n}(x) \times \partial_\nu \bm{n}(x) ] .
\end{align}

The remaining issue is to answer how to define and obtain the color unit vector field $\bm{n}(x)$ from the original Yang-Mills theory written in terms of $\mathscr{A}_\mu(x)$ alone. 
A procedure has been given in our reformulation of Yang-Mills theory in the continuum spacetime.\cite{KMS05,KMS06}
By introducing $\bm{n}(x)$ field in addition to $\mathscr{A}_\mu(x)$, we have a gauge theory with the enlarged gauge symmetry $\tilde{G}$, called the master Yang-Mills theory. 
We propose a constraint\cite{KMS06}(called the new Maximal Abelian gauge, nMAG) by which 
$\tilde{G}:=SU(2)_{\omega} \times [SU(2)/U(1)]_{\theta}$ in the master Yang-Mills theory is broken down to the diagonal subgroup: $G'=SU(2)$: 
Minimize the functional $\int d^D x \frac12 g^2 \mathbb X_\mu^2$ w.r.t.  the enlarged gauge transformations: 
\begin{align}
 0 = \delta_{\omega, \theta} \int d^D x \frac12 g^2 \mathbb X_\mu^2 
= \delta_{\omega, \theta} \int d^D x   (D_\mu[\mathscr{A}]{\bm n})^2 .
 \label{MAGcond}
\end{align}
It has been shown that nMAG determines the color field $\bm{n}(x)$ as a functional of a given configuration of $\mathscr{A}_\mu(x)$. 
Therefore, if we impose the new MAG (\ref{MAGcond}) to the master-Yang--Mills theory,
we have a gauge theory (called the Yang--Mills theory II) with the local gauge symmetry 
$G':=SU(2)_{local}^{\omega'}$ with  
$\bm\omega'(x)=(\bm\omega_\parallel(x),\bm\omega_\perp(x)=\bm\theta_\perp(x))$, 
which is a diagonal SU(2) part of the original 
$\tilde{G}:=SU(2)_{local}^{\omega} \times [SU(2)/U(1)]_{local}^{\theta}$.  
The local gauge symmetry $G'$  of the Yang--Mills theory II for new variables is  
\begin{align}
  \delta_{\omega'} \bm{n}  =& g \bm{n} \times \bm{\omega'}  ,
\\
 \delta_{\omega'} c_\mu =&    \bm{n} \cdot \partial_\mu \bm{\omega'}   ,
\\
  \delta_{\omega'}  \mathbb{X}_\mu =&  g \mathbb{X}_\mu \times \bm{\omega'} ,
\\
\Longrightarrow  & 
\delta_{\omega'}  \mathbb{V}_\mu =   D_\mu[\mathbb{V}] \bm{\omega'}   , 
\quad
\delta_{\omega'}  \mathscr{A}_\mu  
=   D_\mu[\mathscr{A}] \bm{\omega'}  .
\end{align}
This should be compared with the conventional MAG.
The old MAG leaves local $U(1)$ and global U(1) unbroken, but breaks global SU(2).
The new MAG leaves local G'=SU(2) and global SU(2) unbroken (color rotation invariant).  This is an advantage of this reformulation.

\section{Gauge-invariant gluon mass and infrared Abelian dominance}

Note that the mass dimension-2 operator $\mathbb X_\mu^2$ is invariant under the local gauge transformation II:
\begin{align}
 \delta_\omega' \mathbb{X}_\mu^2(x) =   0 .
\end{align}
We claim the existence of gauge-invariant dimension-two condensate\cite{Kondo06}
\begin{align}
 \langle 0| \mathbb{X}_\mu(x) \cdot \mathbb{X}_\mu(x) |0  \rangle_{\rm YM} \ne 0 .
\end{align}
How does such a {gauge-invariant} dimension-two condensates 
occur? 
It can be induced from the self-interactions among gluons:
\begin{align}
- \frac{1}{4} (g \mathbb{X}_\mu \times \mathbb{X}_\nu )^2 , 
\end{align}
since a rough observation yields 
\begin{align}
 & -\frac{1}{4}(g \mathbb{X}_\mu \times \mathbb{X}_\nu) \cdot (g \mathbb{X}^\mu \times \mathbb{X}^\nu) 
 \nonumber\\
\to & \frac{1}{2}g^2 \mathbb{X}^A_\mu \left[\left\langle -\mathbb{X}^2_\rho \right\rangle \delta^{AB} - \left\langle -\mathbb{X}^A_\rho \mathbb{X}^B_\rho \right\rangle \right] \mathbb{X}^B_\mu 
=   \frac{1}{2} M_X^2 \mathbb{X}_\mu \cdot \mathbb{X}_\mu ,
\quad M_X^2 = \frac23 g^2 \left\langle -\mathbb{X}^2_\rho \right\rangle .
\label{mass-term}
\end{align} 
This idea can be made precise. 
See ref.\cite{Kondo06} for details.
Consequently, the gauge-invariant mass term for "off-diagonal" $\mathbb{X}_\mu$ gluons is generated. 
This mass term is invariant under the local SU(2) gauge transformation II.
Then the $\mathbb{X}_\mu$ gluons to be decoupled in the low-energy (or long-distance) region.  
This leads to the {infrared Abelian dominance}: 
The Wilson loop average is entirely estimated by the field $\mathbb{V}_\mu$ alone.

\section{Lattice formulation and numerical simulations}

Our reformulation of Yang-Mills theory in the continuum spacetime \cite{KMS06,KMS05} has been implemented on a lattice to perform numerical simulations as follows.

\noindent
$\bullet$ Non-compact formulation\cite{KKMSSI06} has lead to

$\cdot$ generation of color field configuration $\bm{n}(x)$ 

$\cdot$ restoration of color symmetry (global gauge symmetry)
 
$\cdot$ gauge-invariant definition of magnetic monopole charge

\noindent
$\bullet$ Compact formulation I \cite{IKKMSS06} has succeeded to show 

$\cdot$ magnetic charge quantization subject to the Dirac   quantization condition $gg_m/(4\pi) \in \mathbb{Z}$ 

$\cdot$ magnetic monopole dominance in the string tension 

\noindent
$\bullet$ Compact formulation II \cite{SIKKMS06} has confirmed

$\cdot$ the non-vanishing gluon mass $M_X=1.2{\rm GeV}$
(cf. MAG result\cite{AS99})

\section{Conclusion and discussion}

Using a nonlinear change of variables, we have succeeded to separate the original SU(2) gluon field variables $\mathscr{A}_\mu$ into  ``Abelian'' part $\mathbb{V}_\mu$ and the ``remaining'' part $\mathbb{X}_\mu$ {\it without breaking color symmetry}:
$$
 \mathscr{A}_\mu = \mathbb{V}_\mu + \mathbb{X}_\mu ,
$$
in the following sense. 

\noindent
$\bullet$ $\mathbb{V}_\mu$ are responsible for quark confinement:
the DP version of the non-Abelian Stokes theorem tells us that the non-Abelian Wilson loop operator is entirely rewritten in terms of the SU(2) invariant {\it Abelian} field strength $G_{\mu\nu}$ defined from the variable $\mathbb{V}_\mu$.  
  This specification leads to a definition of gauge-invariant magnetic monopoles with the magnetic charge subject to Dirac quantization  condition
(which is confirmed by analytical and numerical methods)
and magnetic monopole dominance in the string tension (confirmed by numerical method).

\noindent
$\bullet$ $\mathbb{X}_\mu$ could decouple in the low-energy regime: 
This is because the $\mathbb{X}_\mu$ gluon  acquires the gauge-invariant mass dynamically through the non-vanishing vacuum condensation of mass dimension--two $\left< \mathbb{X}_\mu^2 \right> \ne 0$.  
This leads to the infrared ``Abelian'' dominance.
We have examined a possibility of dimension--two condensate $\left< \mathbb{X}_\mu^2 \right> \ne 0$ by analytical\cite{Kondo06}  and numerical\cite{KKMSS05} methods.

In ref.\cite{KOSSM06}, we have discussed how the Faddeev model can be regarded as a low-energy effective theory of Yang-Mills theory to see the mass gap.

\end{document}